\documentclass[twocolumn]{aastex62}

\graphicspath{{./}{figures/}}

\accepted{March 31, 2019}

\shorttitle{A strategy for LSST to unveil a population of kilonovae without GW triggers}
\shortauthors{Andreoni et al.}

\usepackage{multirow}

\usepackage{graphicx}	
\usepackage{amsmath}	
\usepackage{amssymb}	
\usepackage{soul} 

\begin{document}

\title{A strategy for LSST to unveil a population of kilonovae without gravitational-wave triggers}

\correspondingauthor{Igor Andreoni}
\email{andreoni@caltech.edu}

\author[0000-0002-8977-1498]{Igor Andreoni}
\affil{Division of Physics, Mathematics and Astronomy, California Institute of Technology, \\
1200 E California Blvd, Pasadena, CA 91125, USA }

\author{Shreya Anand}
\affil{Division of Physics, Mathematics and Astronomy, California Institute of Technology, \\
1200 E California Blvd, Pasadena, CA 91125, USA }

\author{Federica B. Bianco}
\affil{{Department of Physics and Astronomy, University of Delaware, Newark, DE, 19716, USA
2}}
\affil{{Joseph R. Biden, Jr,. School of Public Policy and Administration, University of Delaware, Newark, DE, 19716, USA
2}}
\affil{{Data Science Institute, University of Delaware, Newark, DE, 19716, USA
2}}
\affil{{Center for Urban Science and Progress, New York University, 370 Jay St,
Brooklyn, NY 11201, USA}}

\author{S. Bradley Cenko}
\affil{Astrophysics Science Division, NASA Goddard Space Flight Center, MC 661, Greenbelt, MD 20771, USA}
\affil{Joint Space-Science Institute, University of Maryland, College Park, MD 20742, USA}

\author{Philip S. Cowperthwaite}
\altaffiliation{NASA Hubble Fellow}
\affil{The Observatories of the Carnegie Institution for Science, 813 Santa Barbara St., Pasadena, CA 91101, USA}

\author[0000-0002-8262-2924]{Michael W. Coughlin}
\affil{Division of Physics, Mathematics and Astronomy, California Institute of Technology, \\
1200 E California Blvd, Pasadena, CA 91125, USA }

\author{Maria Drout}
\affil{Department of Astronomy and Astrophysics, University of Toronto, 50 St. George Street, \\ Toronto, Ontario, M5S 3H4 Canada}
\affil{The Observatories of the Carnegie Institution for Science, 813 Santa Barbara St., Pasadena, CA 91101, USA}

\author[0000-0001-8205-2506]{V. Zach Golkhou} 
\altaffiliation{Moore-Sloan and DIRAC Fellow}
\affil{DIRAC Institute, Department of Astronomy, University of Washington, 3910 15th Avenue NE, Seattle, WA 98195, USA} 
\affil{The eScience Institute, University of Washington, Seattle, WA 98195, USA}

\author[0000-0001-6295-2881]{David L.\ Kaplan}
\affil{Department of Physics, University of Wisconsin-Milwaukee, PO Box 413, Milwaukee, WI, 53201, USA} 

\author{Kunal P. Mooley}
\affil{Division of Physics, Mathematics and Astronomy, California Institute of Technology, \\
1200 E California Blvd, Pasadena, CA 91125, USA }

\author{Tyler A. Pritchard}
\affil{Center for Cosmology and Particle Physics, New York University, 726 Broadway, New York, NY 10004, USA}

\author{Leo P. Singer}
\affil{Astrophysics Science Division, NASA Goddard Space Flight Center, MC 661, Greenbelt, MD 20771, USA}
\affil{Joint Space-Science Institute, University of Maryland, College Park, MD 20742, USA}

\author[0000-0003-2601-1472]{Sara Webb}
\affil{Centre for Astrophysics and Supercomputing, Swinburne University of Technology,\\ Mail Number H29, PO Box 218, 31122, Hawthorn, VIC, Australia  }
\affil{ARC Centre of Excellence for Gravitational Wave Discovery (OzGrav), Australia}

\collaboration{with the support of the LSST Transient and Variable Stars Collaboration}



\begin{abstract}

We present a cadence optimization strategy to unveil a large population of kilonovae using optical imaging alone.  These transients are generated during binary neutron star and potentially neutron star--black hole mergers and are electromagnetic counterparts to gravitational-wave signals detectable in nearby events with Advanced LIGO, Advanced Virgo, and other interferometers that will come online in the near future. Discovering a large population of kilonovae will allow us to determine how heavy element production varies with the intrinsic parameters of the merger and across cosmic time. The rate of binary neutron star mergers is still uncertain, but only few ($\lesssim 15$) events with associated kilonovae may be detectable per year within the horizon of next-generation ground-based interferometers.  
The rapid evolution ($\sim$ days) at optical/infrared wavelengths, relatively low luminosity, and the low volumetric rate of kilonovae makes their discovery difficult, especially during blind surveys of the sky. We propose future large surveys to adopt a rolling cadence in which $g$-$i$ observations are taken nightly for blocks of 10 consecutive nights.  With the current baseline2018a cadence designed for the Large Synoptic Survey Telescope (LSST), $\lesssim 7.5$ poorly-sampled kilonovae are expected to be detected in both the Wide Fast Deep (WFD) and Deep Drilling Fields (DDF) surveys per year, under optimistic assumptions on their rate, duration, and luminosity. 
We estimate the proposed strategy to return up to $\sim 272$ GW170817-like kilonovae throughout the LSST WFD survey, discovered independently from gravitational-wave triggers. 

\end{abstract}


\section{Introduction}
\label{sec: introduction}

Binary neutron star mergers have long been predicted to be associated with short gamma-ray bursts \citep[sGRBs][]{Blinnikov1984,Paczynski1986,Eichler1989a, Narayan1992,Fong&Berger2013} and optical/near-infrared transients called kilonovae (KN) or macronovae \citep[see for example][]{Li1998,Kulkarni2005,Rosswog2005}.  A ``living review" of KNe can be found in \cite{Metzger2017b}.

The low electron fraction of neutron star merger ejecta favors the production of heavy elements such as lanthanides and actinides via rapid neutron capture (r-process).  The optical and infrared transient is powered by the decay of these unstable nuclei.
The nature of the emission is governed by the neutron-richness and production of lanthanides, or lack thereof.  ``Red" KNe are usually associated with lanthanide-rich, mildly relativistic dynamical ejecta present in the equatorial plane tidal tails.  Lanthanide-rich material increases the opacity, causing the emission to become brighter in the infrared \citep[e.g.,][]{Hotokezaka2013, Barnes2013,Kasen2013,Tanaka2013,Tanaka2018}. 
A ``blue" KN component can be produced in lanthanide-poor ejecta present in polar disk winds, generated from the accretion of a disk of material onto the merger remnant \citep[e.g.,][]{Metzger2010a,Roberts2011}.  Such a blue component may be visible only under certain viewing angle conditions \citep{Kasen2017Nat}.

\begin{figure*}
    \centering
    \includegraphics[width=0.9\columnwidth]{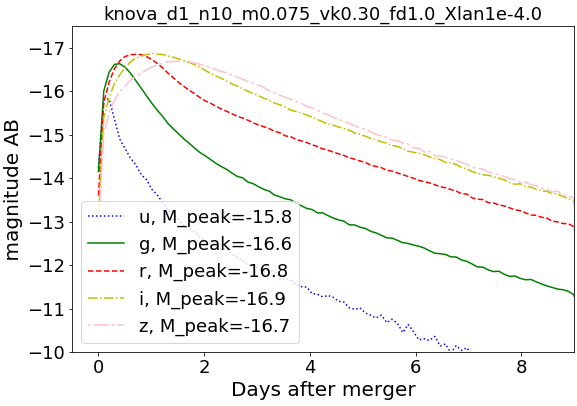}
    \includegraphics[width=0.9\columnwidth]{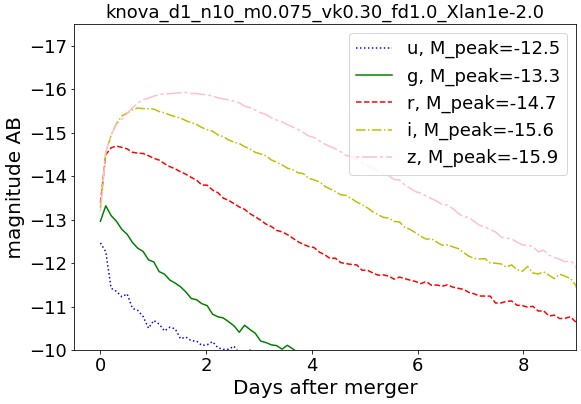}
    \includegraphics[width=0.9\columnwidth]{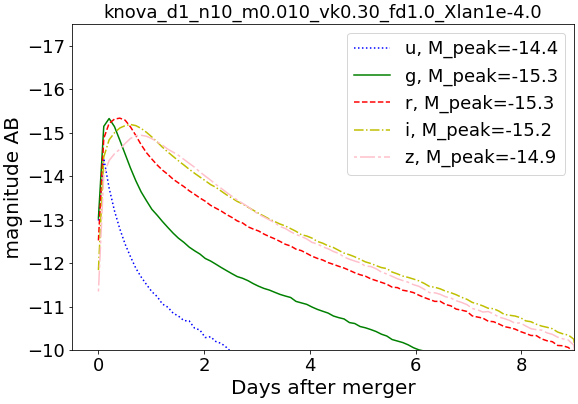}
    \includegraphics[width=0.9\columnwidth]{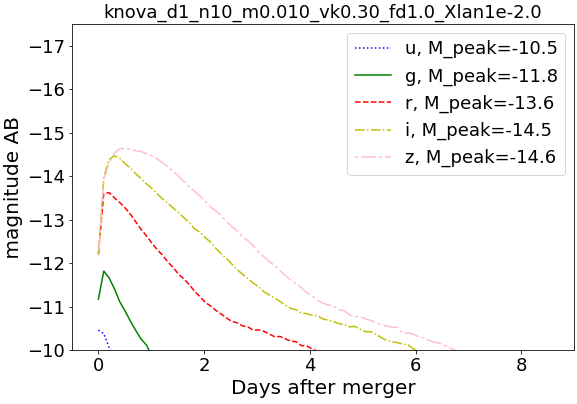}
    \includegraphics[width=0.9\columnwidth]{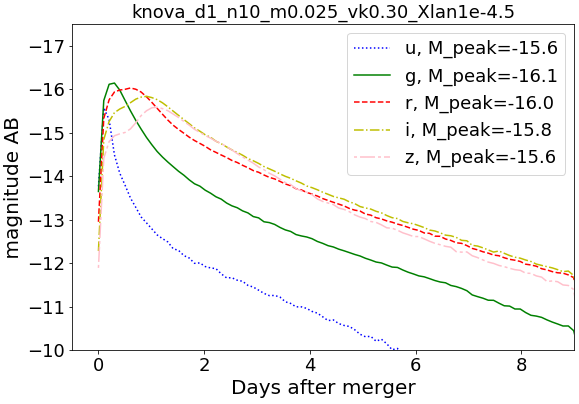}
    \caption{KN models from \cite{Kasen2017Nat} convolved with Dark Energy Survey filters, similar to LSST filters. The two plots at the top show KNe with high ejecta mass (0.075$\textup{M}_\odot$) and the two central panels show KNe with low ejecta mass (0.01$\textup{M}_\odot$). The plots on the left are created assuming low lanthanide mass fraction (Xlan=$10^{-4}$) and those of the right assuming high lanthanide mass fraction (Xlan=$10^{-2}$).  The bottom plot shows model light curves built with parameters that best fit the KN counterpart to GW170817 \citep[0.025$\textup{M}_\odot$, Xlan=$10^{-4.5}$;][]{Arcavi2018}.}
    \label{fig:kasenmodels}
\end{figure*}

Signatures of KNe were first found during the follow-up of sGRBs \citep{Perley2009,Tanvir2013a,Berger2013k,Gao2015,Jin2015,Jin2016,Troja2018KN,Jin2019arXiv}, but never found thereafter during blind surveys of the sky \citep[e.g.,][]{Doctor2017,Scolnic2018}, possibly due to low KN luminosity and rapid evolution (Figure\,\ref{fig:kasenmodels}-\ref{fig: KN colors}).  

On August 17, 2017 a KN was discovered in association with the gravitational wave (GW) signal of a binary neutron star merger, known as GW170817 \citep{Coulter2017,Valenti2017,Arcavi2017GW,Tanvir2017,Lipunov2017,Soares-Santos2017}. The identification of the KN was key to pin-point the localization of the merger and to prompt multi-wavelength follow-up \citep[see][and references therein]{Abbott2017MMA}. Optical and infrared observations \citep{Andreoni2017gw, Arcavi2017GW, Chornock2017, Coulter2017, Cowperthwaite2017, Diaz2017, Drout2017Sci, Evans2017Sci, Hu2017, Kasliwal2017, Lipunov2017, McCully2017, Pian2017Nat, Smartt2017, Tanvir2017, Shappee2017, Utsumi2017} showed that the transient reddened and faded away more rapidly than other known transients, with $\Delta g > 5$ and $\Delta$T$\gtrsim 6000$\,K in the first week of observations \citep[see also][]{Kilpatrick2017Sci, Nicholl2017, Siebert2017, Villar2017}.  
Detailed modeling of the photometric data showed the light curves were consistent with r-process heating, where a blue and a red component are generated from lanthanide-poor polar winds and lanthanide-rich material, respectively. Models with ejecta with mass $\sim$0.05\,M$_\odot$ and 0.1-0.3$c$ velocity fit the data well, where the blue component is roughly half the mass and three times the velocity of the red component \citep{Kasen2017Nat}.  The presence of a third emission component was also considered \citep{Cowperthwaite2017, Kilpatrick2017Sci, Tanaka2017}.   
Some works specifically addressed the evidence for r-process nucleosynthesis following the merger, including \cite{Cowperthwaite2017, Chornock2017, Drout2017Sci, Smartt2017, Pian2017Nat, Kasen2017Nat, Kasliwal2017} and indicated neutron star mergers may be dominant sites for heavy-element production in the Universe \citep[e.g.,][]{Rosswog2018}.       
The combined GW and electromagnetic information was used to measure cosmological parameters without tying to a distance ladder \citep{Abbott2017cosmology,Hotokezaka2018arXiv}.

The detection of neutron star-black hole mergers and their electromagnetic counterparts is one of the most exciting challenges for multi-messenger astronomers in the near future. Such mergers 
may also produce KNe \citep{Kasen2017Nat}.  
The KAGRA and LIGO-India interferometers are expected to come online and join Advanced LIGO (AdLIGO) and Advanced Virgo (AdVirgo) for  detecting GW signals between 2020--2025 \citep{Abbott2018prospects}. A growing network of GW observatories will provide more and better-localized GW detections.  With the addition of the LIGO-India and KAGRA detectors to the AdLIGO-AdVirgo network at its design sensitivity, the number of binary neutron stars detected per year is estimated to increase by a factor of 2-3, though the BNS range for KAGRA and AdVirgo will be lower than the expected 190 Mpc range for the remaining three detectors \citep{Abbott2018prospects}. The median 90\% credible localization region for binary neutron star mergers is expected to be reduced from $\sim 150$\,deg$^2$ (AdLIGO and AdVirgo) to $\sim 10$\,deg$^2$ when LIGO-India and KAGRA join the AdLIGO and AdVirgo network.  Consequently, the follow-up of these GW triggers should lead to the detection of a sample of KNe.  However, KN discoveries un-triggered by GW detection is likely going to play an important role in complete population studies, especially if the GW detectors yield fewer binary neutron star events than expected.  

Rates of binary neutron star mergers are still highly uncertain. \cite{Abbott2017GW170817discovery} estimate a binary neutron star merger rate R=$1540^{+3200}_{-1200}$ Gpc$^{-3}$\,yr$^{-1}$ based solely on GW searches.  However, the lack of KN detection in past optical surveys constrains this rate toward its lower limit, placing 3$\sigma$ upper limits to R $< 800$\,Gpc$^{-3}$yr$^{-1}$ for GW170817-like events and a more conservative value of R $< 1600$\,Gpc$^{-3}$yr$^{-1}$ assuming that the typical KN is 50\% as luminous as the GW170817 KN \citep{Kasliwal2017}. Assuming a upper limit R $< 800$\,Gpc$^{-3}$yr$^{-1}$, a total number of $\lesssim 23$\,y$^{-1}$ binary neutron star mergers can be expected to occur within the nominal AdLIGO horizon of 190\,Mpc, only $\lesssim 15$ of which will be detectable due to the non-homogeneous antenna pattern of GW detectors and their duty cycle.  Moreover, only a fraction of those could have a detectable KN, as some could be located too close to the Sun to observe with optical telescopes, or generate faint transients beyond the detection limit of available telescopes. Obscuration and confusion by the Galactic plane is also a significant limitation to the detection of counterparts. 
In light of these limitations, the expected binary neutron star detection rate of 4-80\,events yr$^{-1}$ for the AdLIGO-AdVirgo-KAGRA network after 2020, based only on GW searches \citep{Abbott2018prospects}, will likely provide only a few tens of triggers throughout the next decade. 

In the next few years, the new Large Synoptic Survey Telescope \citep[LSST,][]{Ivezic2008} is expected to be a game-changing facility in astrophysics.   Time-domain astronomy will particularly benefit from the large $\sim 10$\,deg$^2$ field of view (FoV) of the camera combined with the depth achievable with the 8.4\,m-diameter primary mirror.  
The observing strategy proposed in this work is particularly suitable for optimizing the main Wide Fast Deep survey of LSST (see Section\,\ref{sec: LSST current}).

This paper is organized as follows.  Section\,\ref{sec: importance KN detections} expands on the reasons why discovering KNe independently from GW triggers can be important enough to deserve designing new observing strategies with a major facility such as LSST.  In Section\,\ref{sec: LSST current} we describe the current plans for observations with LSST and expectations for serendipitous KN discoveries. We present the design of a new strategy optimized for KN discovery in Section\,\ref{sec: proposed strategy} and we quantify its performance in Section\,\ref{sec: performance}. The proposed strategy is then discussed in Section\,\ref{sec: discussion} and benefits for other science cases are presented in Section\,\ref{sec: other science cases}. 

\section{Importance of finding kilonovae without gravitational-wave triggers}
\label{sec: importance KN detections}

Determining robust rates of KNe is of primary importance because it directly constrains the rate of binary neutron star mergers. Hundreds of KN detections would allow us to understand the distribution of parameters such as ejecta mass, ejecta velocity, and opacity, along with a better understanding of the jet physics (when combined with high-energy observations) and the dependence of observed properties on the viewing angle.  Moreover, a better knowledge of KN rates and properties could shed some light on the current debate on whether KNe \citep[e.g.,][]{Kasen2017Nat} or collapsing massive stars \citep{Siegel2018arXiv,Siegel2019arXiv} are the dominant sites for heavy-element nucleosynthesis in the Universe.
Furthermore, searches for distant events can allow us to understand how the KN rate varies with redshift.  A blind survey for fast transients with well planned color measurements can help us understand the distribution of their properties more completely, highlighting which different types of KNe exist and how ``typical'' or atypical GW170817 was.

KNe discovered with LSST should provide a large sample of host galaxies, allowing us to search for correlations between KN properties and galaxy morphology, star formation, and metallicity \citep{Levan2017,Pan2017,Im2017ApJ, Blanchard2017ApJ, Andreoni2017gw}.  Even if spectroscopic measurements of the active transients are not performed, large telescopes can provide high-quality observations of their hosts. 
The detection of KNe with LSST could also trigger reverse-searches for low-significance signatures in data acquired with GW and neutrino detectors \citep{Acernese2007}. Such an approach can be valuable
especially when only one detector is online and accurate GW localization is not possible, even if the significance of the GW detection is high. 

Target of Opportunity (ToO) observations with LSST will serve to search for counterparts shortly after GW triggers are issued \citep{Margutti2018WP,Cowperthwaite2018arXiv}.
Limitations for KNe detection via ToO follow-up include GW detectors experiencing significant downtime, with an average duty cycle of 60\,\%-70\,\% in addition to months in which the interferometers are offline to be upgraded. When one or more interferometer is offline, the whole GW detector network decreases in sensitivity, and the source localization becomes poorer, making the discovery of possible electromagnetic counterparts more difficult.  As in the case of GW170817, KNe can also be found to be associated with sGRBs \citep{Perley2009,Tanvir2013a,Berger2013k,Gao2015,Jin2015,Jin2016,Troja2018KN,Jin2019arXiv} detected with gamma-ray telescopes such as {\it Fermi} or {\it Neil Gehrels Swift Observatory}.  However, sGRBs are typically located at large distances and their relativistic emission is observable only at favorable viewing angles, in contrast to KNe, which are isotropic in their emission and detectable even at low redshifts \citep{Metzger2017b}.  Other limiting factors for GW follow-up include observational constraints that can prevent the discovery of KNe, such as the merger occurring at coordinates too close to the Sun, near the Galactic plane, or in highly dust-obscured regions of their host galaxies.  We note that, in these cases, counterparts to GW events could still be discovered with radio telescopes.  

\section{Expectations for KN discovery with the LSST 2018 baseline cadence}
\label{sec: LSST current}

The overall LSST observing stategy is designed to achieve four main science goals: taking an inventory of the solar system, mapping the Milky Way via resolved stellar population, exploring the transient optical sky, and probing dark energy and dark matter \citep{Ivezic2008}. Using these themes, the original observing strategy as outlined in the LSST Science Book designated roughly 80-90\% of the total observing time to a universal Wide-Fast-Deep (WFD) program, and allotted 10\% of time for Deep Drilling Fields (DDF) and 1\% of time for \emph{mini surveys} for specialized science goals \citep{LSST2009}. The current {\tt baseline2018a}\footnote{\url{http://ls.st/Document-28453}} simulation suggests that the main WFD survey will require 86\% of the total survey time, for a total of 2,372,700 visits. This strategy yields 2293 fields observed in 6 filters over the 10 year survey with a median number of visits as $u$: 61 $g$: 85 $r$: 194 $i$: 193 $z$: 180 $y$: 179. With the presently planned LSST WFD survey strategy, it is expected that only 69 KNe events will be detected over the 10 year lifetime of the survey, with only 5.5 more detections from DDF \citep{Scolnic2018}.  \cite{Scolnic2018} define a few specific criteria for detection, requiring observations of KNe with S/N $>$ 5 in at least two filters within a time-window of 25 days, and for an observation within 20 days before, and within 20 days after the S/N$>$5 observation.

The planned LSST {\tt baseline2018a} cadence strategy is expected to yield $<10$ events per year under optimistic conditions, because the strategy is not optimized for KN discovery \citep{Scolnic2018, Cowperthwaite2018arXiv}.
Only a small number ($<$1\,y$^{-1}$) of such rare and faint transients will be detectable in LSST DDF because of the limited sky area explored \citep{Scolnic2018}.

\begin{figure}
    \centering
    \includegraphics[width=1.\columnwidth]{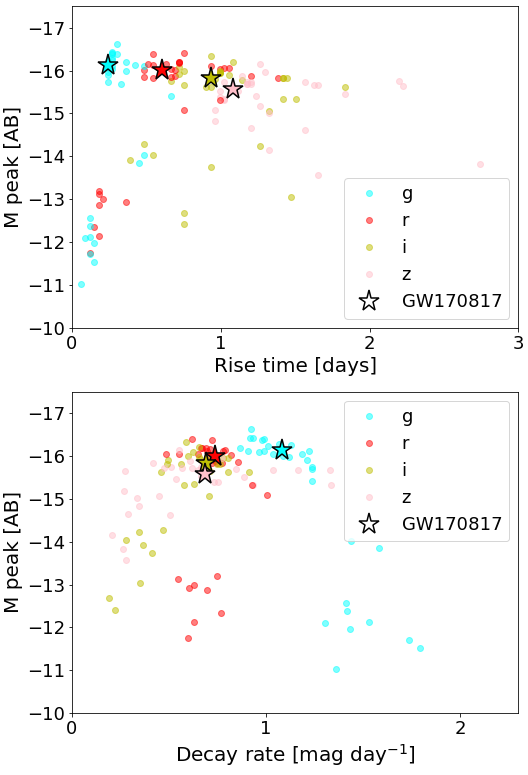}
    \caption{The peak magnitudes of a grid of KN models spanning the space of merger parameters \citep{Kasen2017Nat} are plotted against the rise time ({\it left}) and the fading rate ({\it right}) expected for each model in the first 3 days after peak. Star-shaped markers indicate the model of the grid that best represents GW170817.
Nightly measurements of $g$-$i$ or $i$-$z$ colors can help identify KN candidates and study their properties. }
    \label{fig: KN colors}
\end{figure}

\section{Nightly cadence with 2 filters}  
\label{sec: proposed strategy}

We designed a generic observing strategy aiming at maximizing KN discovery via nightly observations of large regions of sky with at least 2 filters, preferably $g$ and $i$. The strategy is described in Section\,\ref{subsec: strategy description} and its application to LSST is presented in Section\,\ref{subsec: strategy for LSST}.

\begin{figure}
    \centering
    \includegraphics[width=1.\columnwidth]{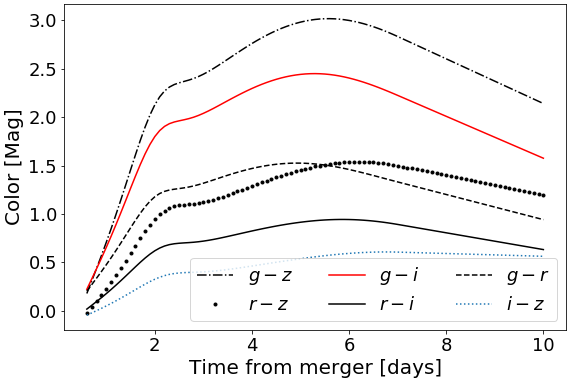}
    \caption{Color evolution of GW170817 in several filter combinations, obtained using the model that \cite{Cowperthwaite2017, Villar2017} developed to describe the GW170817 kilonova. The combination of $g$ and $z$ observations returns the largest magnitude difference, with the steepest rise in the first few days after the merger.  The $gi$ combination, chosen for the strategy recommended in this paper, returns a smaller change than $gz$, but it outperforms other filter combinations. }
    \label{fig: GW170817 color evolution}
\end{figure}

\begin{figure}
    \centering
    \includegraphics[width=1.\columnwidth]{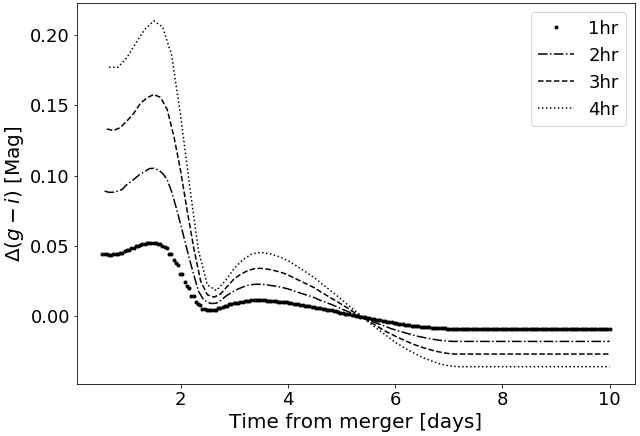}
    \caption{Difference in $g-i$ color assuming a time gap between $g$ and $i$ observations of from 1\,hr to 4\,hr.  Simultaneous $g$ and $i$ observations would lead to $\Delta (g-i) \sim 0$. Time gaps shorter than 1\,hr are preferred, although large gaps of $\sim$4\,hr would lead to up to $\sim 0.2$\,magnitude error in color in the first couple of days after the merger, $<0.05$\,magnitude error at later phases. Such differences may be comparable with -- or smaller than -- photometric uncertainties.  
    }
    \label{fig: color times GW170817}
\end{figure}

\subsection{Strategy description}
\label{subsec: strategy description}

The KN associated with GW170817 faded in $g$ faster (almost 2 mag in 24 hours, Figure\,\ref{fig: KN colors}) and reddened faster (from $g-z=-0.3$ to $+1.3$ in 24 hours) than any other known or theorized transient. Such rapid evolution in brightness and color can be used to identify KN candidates. 

Using $g$-$z$ pairs of observations would be preferred in order to measure more dramatic color changes (Figure\,\ref{fig: GW170817 color evolution}) and to target redder KNe. 
However, our calculations show that fewer KNe are expected to be detected using $z$ filter than using the adiacent and more sensitive $i$ filter, for all types of KN models within the range of ejecta masses and lanthanide fraction that we considered (see Section\,\ref{sec: performance} and Table\,\ref{table: KN numbers}).  Figure\,\ref{fig: GW170817 color evolution} shows that the $gi$ combination provides larger and more rapidly evolving color changes than other filter combinations, other than $gz$, for GW170817.  Here the KN is described by the model developed by \cite{Cowperthwaite2017, Villar2017}. \cite{Bianco2019PC} obtain that $gi$ pairs are among the preferred options when applying machine-learning techniques to choose which filter combinations can best separate fast transients from Type Ia supernovae and other slowly-evolving transients using intra-night observations.

Observations in $g$ and $i$ filters should be performed as close in time as possible to provide reliable color information of fast-evolving transients.  A time lag of $\gtrsim 30$\,min is usually adopted to flag moving objects.  Figure\,\ref{fig: color times GW170817} represents the difference in color of GW170817 when considering observing gaps between $g$ and $i$ observations of 1\,hr, 2\,hr, 3\,hr, and 4\,hr. Within this time gap range, the error in $g-i$ color measurement is $\lesssim 0.20$ magnitudes at all phases of the KN, decreasing to $\lesssim 0.05$\,magnitudes about 2\,d after the merger. Such errors may lie within the uncertainty of photometric measurements, in which case they would affect the derivation of other quantities in a negligible way. 

The choice of observing for 10 consecutive nights is dictated by several factors:
\begin{itemize}
    \item KN models \citep{Kasen2017Nat} are expected to fully evolve in $\lesssim$10 days, so high cadence is necessary for their discovery. A blue component generated from lanthanide-poor disk winds can be present and is expected to be visible for only 3-4 days. A compilation of light curves estimated for a range of ejecta masses and velocities are presented in Figure\,\ref{fig:kasenmodels}.     
    \item Ten consecutive nights of observations would allow us to obtain well-sampled light curves. At least 2-3 data points per band combined with upper limits before and/or after the detection can help uniquely identify KNe, reducing the number of contaminant sources (see Section\,\ref{sec: discussion}).

    \item A sequence of 10 consecutive nights make it easier to schedule observations in dark and grey time to maximize the depth of the observations, and therefore the number of discoveries. 
    
\end{itemize}

\subsection{A new strategy for LSST}
\label{subsec: strategy for LSST}

Specifically in case of LSST, we propose to adopt a ``rolling" cadence for the WFD survey in which, every year, $gi$ observations are taken every night for 10 consecutive nights over a sky area of 500-700\,deg$^2$, considering a total sky area of 18,000\,deg$^2$ to be covered by LSST as planned in the current strategy. 
The cadence that we propose is organized in two possible ways:

\begin{itemize}
    \item A) Assuming observations in the $gi$ bands take place for 20 nights a month with the proposed nightly cadence, broken up into two blocks of 10 consecutive nights, the sky area to be observed per set of consecutive nights is approximately 750\,deg$^2$ per night. After 10 nights, a different 750\,deg$^2$ area is observed for the next block of 10 nights. Observing for 20 days per month allows us to avoid bright time observations.
    
    \item B) Assuming observations taking place 360 days per year, the sky area to be observed every block of 10 consecutive nights is 500\,deg$^2$ per night.  After 10 nights, a different 500\,deg$^2$ area is observed for the following block of 10 nights. This strategy is less effective than A) because it includes bright time observations.  
    
\end{itemize}

Such cadence allows nightly observations to be performed on a sky area 
$\sim$12.5 times larger than the one explore with the DDF survey every night.  
Observations in the remaining  $urzy$ filters (and additional 130 $i$-band pointings) are not constrained by this proposal and may be obtained with whatever alternative cadence that will best fulfill the broad LSST science objectives.  These observations will also serve to provide longer baseline for the $gi$ observations, allowing us to further reject supernovae and monitor the long-term behaviour of the discovered transients.

 Every set of 10 consecutive observing nights is independent from each other, so every year the time at which a certain sky area is imaged can be optimized for other science cases, as long as it is imaged for 10 consecutive nights every year. The specific month at which each sky area is covered can differ every year, in order to enable the imaging of 18,000\,deg$^2$ per year. However different options for the amount of total sky area to cover can be considered, as the $\sim$ 18,000\,deg$^2$ assumed here may differ from the final choice of strategy for LSST.  

Moving observations away from the Galactic plane to avoid high-extinction regions is also preferable, leaving specific mini-surveys the task of exploring the Galactic plane with the cadence that they find most appropriate, as recommended by \cite{Lochner2018WP}.

As our proposal addresses the cadence of the original LSST strategy, {\it it should not add any significant overheads in time} to the WFD survey as planned in 2018.  However, more $g$-band observations \cite[at least 100 median visits per pointing, as described in][, in its early 2019 version]{LSST2009} are required to best exploit the potential of the proposed strategy. In general, we suggest that budgeting more $g$-band visits than presently planned for LSST should be considered, in favor of fast-transient astronomy as well as for a broad range of time-domain projects, including those focused on cosmology.
When estimating the approximate total amount of time required for our main proposed strategy, we consider 750\,deg$^2$ tiling per night, in blocks of 10 consecutive nights, using $g$ and $i$ filters. 
Assuming 2 visits (one in $g$ and one in $i$) with 30s exposure time, the required observing time per night is: 

2 visits $\times$ [(30s exposure) + (120s slew) + (5s settle and readout)] $\times$ (750 deg$^2$) / (10 deg$^2$ FoV) = 6.46\,hr 

Therefore the annual time necessary to perform such observations is $\sim 1550$ hours per year, as planned to be allocated for $gi$ observations \citep{LSST2009}.

\section{Performance Evaluation}
\label{sec: performance}

\begin{table*}
\centering
\caption{We explore the detectability of a set of KNe modelled with different ejecta mass and lanthanide mass fraction. First we compute the expected number of KNe recovered with at least one detection in both $g$ and $i$ filters ($gi$) assuming that we can sample the light curve with nightly cadence, therefore covering the kilonova peak at both bands. The ``Raw potential" represents the number of KN that could be detected during LSST WFD survey (at least one data point in at least one filter). We compute the number of KNe that we expect to recover in the survey with at least 2 detections in the same filter with nightly cadence (C1),  2-night cadence (C2), and 3-night cadence (C3).  Numbers are computed considering a KN rate of 1,000\,Gpc$^{-3}$y$^{-1}$.}
\label{table: KN numbers}
\begin{tabular}{lcccccc}
\hline \hline
 KN model & $gi$ & Filter & Raw & C1  &  C2  &  C3  \\ 
&(C1)&& potential &  &   &    \\ 
\hline
\multirow{3}{*}{GW170817} && $g$ & 782 &131 & 42 & 19\\
&272&$i$ & 199 & 82 & 37 & 16 \\
&&$z$ & 21 & 9 & 3 & 1 \\
\hline

\multirow{3}{*}{Low ej. mass, low Xlan} && $g$ & 315 & 49 & 15 & 6 \\
& 96 &$i$ & 87 & 37 & 14 & 6   \\
&&$z$ & 10 & 4 & 1 & 1   \\

\hline
\multirow{3}{*}{Low ej. mass, high Xlan} && $g$ & 3 & 0 & 0 & 0 \\
&1 &$i$ & 27 & 11 & 4 & 2 \\
&&$z$ & 8 & 4 & 1 & 0   \\

\hline
\multirow{3}{*}{High ej. mass, low Xlan} && $g$ & 1,367 &  237 & 77 & 36  \\
& 460&$i$ & 640 & 276 & 120 & 55  \\
&&$z$ & 96 & 45 & 16 & 6  \\

\hline
\multirow{3}{*}{High ej. mass, high Xlan} && $g$ & 19 & 3 & 1 & 0 \\
&5.9 &$i$ & 121 & 50 & 20 & 9  \\
&&$z$ & 40 & 20 & 5 & 3  \\
\hline
\end{tabular}
\end{table*}

We estimate the number of KNe that could be detectable in different scenarios, considering several cadence options (1, 2, and 3 night cadence) and diverse KN models. 
The luminosity and duration of KNe strongly depend on the ejecta mass, kinetic velocity, and electron fraction among other parameters. 
Details on the numerical radioactive-decay-powered models that we explore are described in \cite{Kasen2017Nat}. We consider in particular 5 models: one that well fits the light curve of GW170817, and the other four built with combinations of low/high ejecta mass (m=0.010/0.075\,$\textup{M}_\odot$), and with low/high lanthanide mass fraction (Xlan=$10^{-4}/10^{-2}$). All the considered models assume kinetic velocity $v_k$=0.3\,c, and density profile exponents d=1 (inner) and n=10 (outer). Figure~\ref{fig:kasenmodels} shows the $griz$ light curves corresponding to the chosen models.

Results of our calculations are presented in Table\,\ref{table: KN numbers}, using detection limits of $g$=24.8, $i$=23.8, and $z$=22.5 \citep{LSST2009}. We focus on $g$, $i$, and $z$ observations in order to identify which combination of filters and cadence allows us to discover the most KNe.
Given the dramatic colors and color evolution of KNe (Figure\,\ref{fig:kasenmodels}), combined $gz$ observations would be preferred, but results in Table\,\ref{table: KN numbers} suggest higher chances of discovering KNe by pairing the $g$ and $i$ filters. 

Calculations show that LSST has the potential to detect a large number of KNe (see the ``Raw Potential" column in Table\,\ref{table: KN numbers}), however individual detections cannot be labelled as ``discoveries" in this context.  The need of future analysis to define optimal selection criteria for KN discovery is discussed in Section\,\ref{sec: discussion}.  

Estimates of the number of KNe that could be discovered can be obtained requiring, for example, at least two detections in one filter on one night, or multiple detections with one filter over multiple nights. 
Our strategy achieves an increase in the number of discovered KNe requiring at least one detection in both $g$ and $i$ filters, assuming a nightly cadence (C1) that allows us to cover the peak time of KNe in both filters and well constrains the duration of the transient.  As the cadence moves from nightly to 2 and 3 nights, the number of discovered KNe decreases by one order of magnitude for every type of KN model.  We also note that \cite{Siebert2017} suggested that LSST can find a large number of GW170817-like KNe if a nightly cadence is adopted.

New simulations of our proposed strategy with the LSST {\it Operations Simulator}\footnote{\software{OpSim}, \cite{Delgado2014}} \citep[\textup{OpSim}][]{Delgado2014} would help us compute even more reliable numbers for expected KN discoveries.  We plan on using the transientAsciiMetric\footnote{\small{\url{https://github.com/fedhere/sims_maf_contrib/blob/master/mafContrib/transientAsciiMetric.py}}} to generate light curves and measure the recovery efficiency for KN models and observed events. 
Our estimates do not account for weather and hardware maintenance. Historical data from other Chilean observatories \citep[e.g.][]{Flaugher2015} suggest that $\sim 10\%$ of nights are usually lost due to poor weather.  Based on \textup{OpSim} simulations, we expect that $\sim 82\%$ of the total 3684 nights planned for LSST survey will provide usable data, accounting for weather and instrumental down-time.

\section{Discussion }
\label{sec: discussion}

One of the main problems to be addressed in KN searches not triggered by GW or gamma-ray signals is defining the selection criteria to identify KNe.  Recent works place constraints on the number of detections and the transient duration to distinguish KNe from other ``contaminant" transients such as Type Ia supernovae \citep{Scolnic2018,Setzer2018ArXiv,Cowperthwaite2018arXiv}.  These constraints include color and light-curve evolution information when addressing KN detection.  \cite{Bianco2019PC} show that intra-night observations can provide enough information to separate fast transient candidates (including KNe) from slower transients, so that follow-up with other facilities can be triggered. More challenging is the rejection of different types of fast transients, whose rates and characteristics are still highly uncertain. The confident detection of more KNe during GW trigger follow-ups in the near future would help determine the actual identification efficiency of KNe, providing a clearer view on the diversity and uncertainty in the KN population \citep[see also][]{Rossi2019ArXiv}.  In other words, a better knowledge of KNe may allow us to {\it discover} KNe more confidently based on photometric information, aside from {\it detecting} them in the data stream.  

The observing strategy that we presented in this paper combines high cadence and color information.  Present and future surveys can benefit from observing with such cadence for KN detection as well as for a number of other science cases (see Section\,\ref{sec: other science cases}). In addition to LSST, sensitive wide-field facilities such as Subaru Hyper Suprime-Cam \citep{Miyazaki2012HSC} and the Dark Energy Camera \citep{Flaugher2015} could be suitable for surveying the sky with the proposed cadence.  

A $gi$-nightly cadence provides a uniform dataset that enables regular monitoring of both brightness and color evolution of the sources.  When such a dataset is obtained, fast transients can be identified with several combinations of selection criteria.  Population studies on transients identified with systematic short-timescale observations could lead to i) a better understanding of the distribution of KN parameters, including ejecta mass, velocity, opacity, and heavy-element content, ii) better ways of separating KNe from other types of fast transients, and iii) possible identification of new classes of dim \citep[e.g.,][]{Kasliwal2011} and rapidly evolving transients \citep[e.g.,][]{golkhou14,golkhou15}.

The ability to find KNe based only on survey data is particularly important in case of deep surveys such as LSST, that are expected to yield a large number (up to $\sim$ million) of transient alerts every night.  Observing strategies with time gaps $>$1-2 nights (in the same filter) should rely on rapid follow-up (hours to days) of many candidates for KN confirmation, possibly performed with 8m-class telescopes if candidates are fainter than $\sim$23 magnitudes.  Although the availability of world-class facilities is limited, they are necessary to observe faint transients especially beyond the AdLIGO/AdVirgo horizon of $\sim$190\,Mpc, where we expect the most serendipitous KN discoveries to take place.  One of the strengths of our proposed strategy is that a large sample of KN candidates with well-sampled light curves can be discovered without the need for prompt spectroscopic classification.  At the same time, $gi$ nightly observations provide excellent information to select a few high-priority candidates for spectroscopic follow-up.  

\begin{figure}
    \centering
    \includegraphics[width=1.\columnwidth]{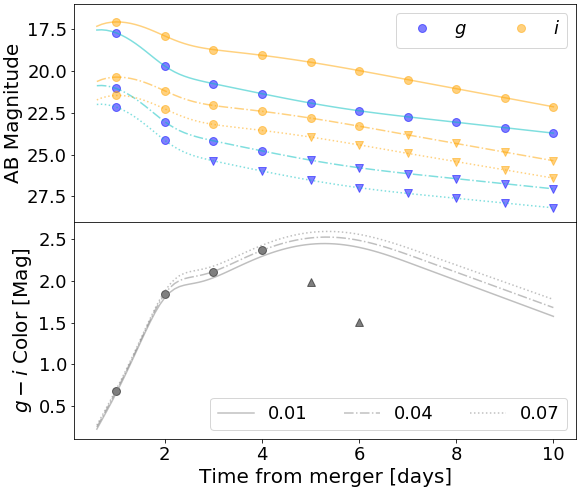}
    \caption{Light curve of GW170817 placed at redshift $z=0.01$ (40\,Mpc, the GW170817 distance), $z=0.04$ (190\,Mpc, the AdLIGO design horizon), and $z=0.07$ (325\,Mpc, A+ LIGO upgrade design horizon). Models are compiled from \cite{Cowperthwaite2017, Villar2017}.   Solid markers indicate the nightly observations that we recommend to perform. Circles mark detection and triangles upper or lower limits. The lower panel shows the $g-i$ color evolution, where detections and lower limits are indicated for redshift $z=0.04$.}
    \label{fig: GW170817 light curve}
\end{figure}

In Figure \ref{fig: GW170817 light curve} we demonstrate how our proposed strategy would sample the lightcurve of a GW170817-like KN placed at 3 different distances, assuming that the observing block starts at the time of the merger.  As expected, an event as close as 40\,Mpc would be fully sampled in both $gi$ bands. Such a nearby object could be detected independently from GW triggers if, for example, GW detectors are offline (see Section\,\ref{sec: importance KN detections}). Light curves are expected to include fewer data points at the AdLIGO design horizon (redshift $z=0.4$) and at the design horizon of the A+ LIGO upgrade ($z=0.07$).  We expect to find more such poorly-sampled events than well-sampled events due to the increase in probed volume close to the detection limit. 

The strategy suggested in this paper could work in good synergy with an intra-night cadence designed for fast transient detection, such as the cadence proposed by \cite{Bianco2019PC}.  The ``{\it Presto-Color}" idea \citep{Bianco2019PC} is based on 3 visits per night in 2 filters ($gi$ or $rz$) and it aims at rapidly recognizing rare fast transients, including KNe.    
If {\it Presto-Color} is adopted with a cadence such as the LSST WFD baseline cadence in 2018, light curves of discovered candidates would be poorly sampled and fast-transient detections would hardly be classifiable, unless successful follow-up observations are performed.  As \cite{Bianco2019PC} suggest, this intra-night cadence is best coupled with a rolling cadence that naturally provides more photometric data points with LSST for those transients recognized to evolve rapidly.  As a result, the strategy that \cite{Bianco2019PC} proposed could be well coupled with the $gi$-nightly cadence described in this work.  

\section{Benefits for other science cases}
\label{sec: other science cases}

The observations proposed herein are fundamentally a combination of three components:  i) intra-night detection and color information, ii) a 10-night high cadence period, observing in two filters every night, and iii) specifically for LSST, a long duration low-cadence period with the remaining four filters $urzy$ and additional $i$ band visits.  This proposed survey strategy is well suited to any object which benefits strongly from intensive observations on 1 - 2 week timescales, while observations outside of this high-cadence window still allow for long-term characterization.  A number of known transient events exist on these timescales whose discovery and characterization to date has been primarily limited by cadence of observations and volume surveyed.  These include, but are not limited to: 
\begin{itemize}
    \item The somewhat enigmatic objects known as ``Fast Blue Optical Transients" or ``Fast Evolving Luminous Transients" \citep[e.g.,][]{Drout2014,Arcavi2016,Rest2018}, that are generally only loosely characterized and whose physical progenitor system and explosion mechanism are poorly constrained.  These objects are often missed in surveys as their rapid rise and evolution on a 3-day cadence often results in only $1-4$ detections depending on brightness, and a larger survey with a daily cadence would allow one to accurately characterize their rise time, light curve shape, and population diversity in an effort to understand their origin.  The intense study and characterization of the extraordinary ultraviolet transient AT2018cow suggests that this class of 'Fast Blue Optical Transients' represents a separate type of astrophysical event, and provides a strong case for their targeted follow-up \citep{Prentice2018, Perley2018, Margutti2019}.
    
    \item Some type Ia supernovae that, in the first $\sim$ 1-3 days after explosion, can show anomalous blue emission, absent in other well-studied events \citep[e.g.,][]{Brown2012}.  This has been inferred to originate from interaction between the SN and the envelope of a companion star \citep[e.g.,][]{Cao2015}. A large sample of Type Ia SN with early color data and a daily cadence would allow the accurate measurement of early blue excess and enable the accurate determination of what fraction of Type Ia SNe come from the single degenerate vs double degenerate progenitor paths.  In addition, it can enable the investigation and modelling of the companion stars while potentially enhancing precision Type Ia cosmology \citep{Jiang2017}.  
    
    \item GRB science through the discovery of an independent sample of optically discovered GRB afterglows \citep[similar to][]{Cenko2015}.  With a completely separate set of systematics, this would enable us to validate our current understanding of long GRBs and allow for a stringent search for true orphan afterglows (GRBs with no detected high-energy emission) in combination with {\it Neil Gehrels Swift Observatory} and {\it Fermi} NASA missions.  It also has the potential to discover a significant population off-axis bursts \citep{Ghirlanda2015}, where the relativistic jet is not directly aligned with the observer but becomes visible as the beaming angle widens.  Some recent studies suggest a significant isotropic emission \citep{Wu2018} may be present which would increase the detection rate.  Such studies would allow us to more tightly constrain the GRB jet opening angle, structure, and environment - particularly when combined with prompt GRB studies \citep[e.g.,][]{golkhou15}.  With these events being rare, and most GRB optical afterglows being detected for less than $\sim 10$ days \citep{Guelbenzu2012}, the high cadence and large area of the LSST WFD survey described here are crucial to discovering a sample large enough to complete these goals.
    \item Core-collapse supernova shock breakout \citep[e.g.,][]{Bersten2018} and shock cooling \citep{De2018Sci} of the extended envelope surrounding the progenitor star.  Those components of supernova explosions evolve on timescales from $\sim$minutes to a few days, and are among the few ways to probe a massive star's stellar structure just prior to its explosion. 
    \item Tidal disruption events (TDEs), for which early multi-band information is key to their recognition \citep[as][stress in their proposal for changes in the LSST observing strategy]{Gezari2018WP,Bricman2018WP}. Same-night color information and nightly cadence can help discover TDEs and trigger follow-up \citep[e.g.,][]{golkhou18}, while the ``third component" of the cadence that we propose (more sparse, low-cadence observations in 4 additional filters, including "u" band which is crucial for TDE characterization) can provide long-term light curves to monitor the evolution of TDEs.     
\end{itemize}

The list above is indicative and not meant to be exhaustive - there are many more studies which would benefit from such a survey strategy. Several works were made public stressing the need for LSST to adopt higher cadence during the main WFD survey \citep{Bianco2019PC,Gezari2018WP,Lochner2018WP} than currently planned.  

The cadence presented in this paper can address any variable object that evolves significantly faster than a Type Ia supernova, or exhibits substantial color evolution. Early, high cadence observations are key for detecting these rare, yet compelling events that are often missed in less frequently cadenced surveys.

\section{Conclusion}

Unveiling a larger population of KNe is necessary to answer questions that could not be answered by ToO observations alone, or by low-cadence, irregular survey strategies. The survey capabilities of the LSST-WFD program offers an unparalleled platform to probe KNe over cosmic time, with the potential to investigate the largest number of detectable KNe in the current age of time-domain astronomy.  
With our proposed $gi$ nightly-cadence strategy hundreds of KNe can be unveiled using LSST, adding $\sim 1$ order of magnitude of events to the number of KNe that are expected to be recovered with the WFD baseline strategy in 2018.  This result would be independent of GW detections of binary neutron star and neutron star--black hole mergers and their follow-up campaigns.

\section*{Acknowledgements}

We thank the anonymous referee for providing feedback that helped us in improving the quality of the paper.

\noindent This work was developed partly within the TVS Science Collaboration and the author acknowledge the support of TVS in the preparation of this paper.

\noindent This work was supported by the GROWTH project funded by the National Science Foundation under Grant No 1545949. GROWTH is a collaborative project between California Institute of Technology (USA), San Diego State University (USA), Los Alamos National Laboratory (USA), University of Maryland College Park (USA), University of Wisconsin Milwaukee (USA), Texas Tech University (USA), University of Washington (USA), Tokyo Institute of Technology (Japan), National Central University (Taiwan), Indian Institute of Astrophysics (India), Indian Institute of Technology (India), Weizmann Institute of Science (Israel), The Oskar Klein Centre at Stockholm University (Sweden), Humboldt University (Germany), Liverpool John Moores University (UK), University of Sydney (Australia).

\noindent
P.S.C. is grateful for support provided by NASA through the NASA Hubble Fellowship grant \#HST-HF2-51404.001-A awarded by the Space Telescope Science Institute, which is operated by the Association of Universities for Research in Astronomy, Inc., for NASA, under contract NAS 5-26555.


\bibliographystyle{aasjournal}
\bibliography{references} 

\end{document}